\documentclass[graybox]{svmult}

\usepackage{makeidx}         % allows index generation
\usepackage{graphicx}        % standard LaTeX graphics tool
\usepackage{multicol}        % used for the two-column index
\usepackage[bottom]{footmisc}% places footnotes at page bottom

\usepackage{newtxtext}       %
\usepackage{newtxmath}       % selects Times Roman as basic font

\makeindex             % used for the subject index
\usepackage[utf8]{inputenc}
    \DeclareUnicodeCharacter{0308}{} % remove Combining Diaeresis character (UTF8-NFD) that breaks things https://unicode-table.com/en/0308/ https://tex.stackexchange.com/q/94418
\usepackage[british]{babel}

\usepackage{amsmath,amsfonts,mathrsfs}
\newcommand*\diff{\mathop{}\!\mathrm{d}}

\usepackage{mathtools}
\usepackage{booktabs}
\usepackage{tabularx}
\usepackage{eurosym}
\usepackage{graphicx}
\usepackage{textcomp}
\usepackage{xcolor}
    \definecolor{green}{rgb}{0.467,0.718,0.337}
    \definecolor{purple}{rgb}{0.588,0.153,0.753}
    \definecolor{lblue}{rgb}{0.353,0.769,0.859}
    \definecolor{orange}{rgb}{0.933,0.627,0.224}
    \definecolor{yellow}{rgb}{0.922,0.769,0.271}
    \definecolor{dred}{rgb}{0.655,0.161,0.129}
\usepackage[hidelinks]{hyperref} % clickable links

\usepackage{cleveref}
    \crefname{algocf}{Algorithm}{Algorithms}
    \crefname{figure}{Fig.}{Figs.}
    \crefname{table}{Table}{Tables}
    \crefname{section}{Section}{Sections}
\usepackage{chemformula}
\usepackage[qualifier-mode=text,detect-mode=true]{siunitx} % number and unit fun
    \DeclareSIUnit{\tCe}{\tonne\of{\ch{CO2}e}}
\usepackage{algorithmic}
\usepackage[linesnumbered,ruled,vlined]{algorithm2e}
    
    \SetCommentSty{mycommfont}
    \newcommand\KwVar[1]{\textit{#1}}
    \newcommand\KwType[1]{{\upshape\texttt{#1}}}
    \SetKw{KwRequire}{require}
    \SetKw{KwAnd}{and}
    \SetKw{KwEmit}{return}
    \SetKwProg{Fn}{Function}{}{end}
    \SetAlCapFnt{\normalfont\small\bfseries} % match svmult.cls, line 1755
    \SetAlCapNameFnt{\normalfont\small} % line 1752
    \SetAlgoCaptionSeparator{\enspace}
    \SetAlCapHSkip{0pt}
\usepackage{float}
\newcommand{\myrule}{\rule[\dimexpr0.5ex - 0.4pt]{1em}{0.8pt}}

\begin{document}

\title*{Carbon Trading with Blockchain}

\author{Andreas Richardson and Jiahua Xu}

\institute{Andreas Richardson \at Imperial College London, London, UK \email{andreas.richardson17@imperial.ac.uk} \at École polytechnique fédérale de Lausanne, Lausanne, Switzerland \\ ORCID: 0000-0002-5466-7279
\and Jiahua Xu \at University College London, London, UK \email{jiahua.xu@ucl.ac.uk} \at École polytechnique fédérale de Lausanne, Lausanne, Switzerland \\ ORCID: 0000-0002-3993-5263}

\authorrunning{A. Richardson and J. Xu}

\maketitle

\abstract{
Blockchain has the potential to accelerate the deployment of emissions trading systems (ETS) worldwide and improve upon the efficiency of existing systems. In this paper, we present a model for a permissioned blockchain implementation based on the successful European Union (EU) ETS and discuss its potential advantages over existing technology. We propose an ETS model that is both backwards compatible and future-proof, characterised by interconnectedness, transparency, tamper-resistance and high liquidity. Further, we identify key challenges to implementation of a blockchain ETS, as well as areas of future work required to enable a fully-decentralised blockchain ETS.
}

\section{Introduction}

Carbon trading systems such as the European Union Emissions Trading System (EU ETS) provide a market mechanism to incentivise emissions reduction, on the basis of \emph{cap and trade}. An overall \emph{cap} on emissions in tonnes of \ch{CO2}-equivalent\footnotemark{} (\si{\tCe}) is imposed, which is translated by a central authority into allowances that are issued to companies. These allowances are surrendered and retired at the end of a reporting period to offset the company's emissions during the period, with the company free to \emph{trade} any surplus allowances on the market \cite{InternationalCarbonActionPartnership2015WhatTrading}. Importantly, should a company have insufficient allowances to cover their (expected) emissions, they are obliged to either purchase surplus allowances from other market participants, or take measures to reduce their emissions. Thus, a high price for allowance units incentivises participants to choose the latter option.

\footnotetext{Scaling factors known as Global Warming Potentials (GWPs) are used to normalise the impact of different Greenhouse Gases (GHGs) emitted.}

On first inspection, this system would seem to be suited to an application of blockchain technology: it involves multiple distributed parties transacting using common currencies, with transactions recorded in an immutable ledger. Indeed, multiple organisations and startups are actively exploring this approach \cite{Baumann2017UsingOutcomes}. However on closer inspection, it is clear that the current state of ETS development poses some critical challenges to the adoption of blockchain technology. For example, one of the frequently-cited advantages of blockchain is the ``disintermediation of trust'' \cite{Braden2019BlockchainSystems,Braden2019BlockchainInstruments,Adams2017TheBlockchain}, meaning a central trusted authority is not required for the network to reach consensus. Yet current ETS designs make heavy use of trusted authorities: a central (governmental) authority is responsible for the distribution of allowances under the cap, whether by direct allocation or through an auction process; further, companies must report their emissions to the central authority and seek verification of this figure from a third-party  \cite{EuropeanCommission2015EUHandbook}. In addition, attacks and fraudulent activities permeating the blockchain space continue to act as a barrier against immediate adoption of the still evolving technology \cite{Braun2019FidentiaX:Blockchain, Xu2019TheScheme}.

As a result, a clear and compelling case must be made to justify the advantages of blockchain over existing technologies. A number of frameworks have been proposed for assessing potential blockchain implementations, considering technical, organisational and legal factors \cite{Kupper2019BlockchainFuture,Braden2019BlockchainInstruments,Braden2019BlockchainSystems}, whilst a series of strategic questions have been raised for business leaders evaluating blockchain's potential \cite{Harbert2019BlockchainChain}. The extreme interest shown in blockchain technology over recent years and the resulting disillusionment with its failure to meet over-hyped promises means the technology is treated with caution; its pros and cons must be carefully weighed \cite{Litan2019Blockchain2019,Litan2019Hype2019,Furlonger2019Hype2019}.

In this paper, we lay out the advantages and challenges of implementing blockchain-backed ETS system, and propose a hybrid model that is both backward compatible and future-proof.
\section{Background}
We first present the EU ETS as a prime example of a contemporary ETS, using it to introduce discussion of weaknesses in current ETS and areas where blockchain technology has strong potential. We additionally present a review of selected literature in this space.

\subsection{EU ETS}
\begin{figure}
    \sidecaption%[t]
    \includegraphics[width=75mm]{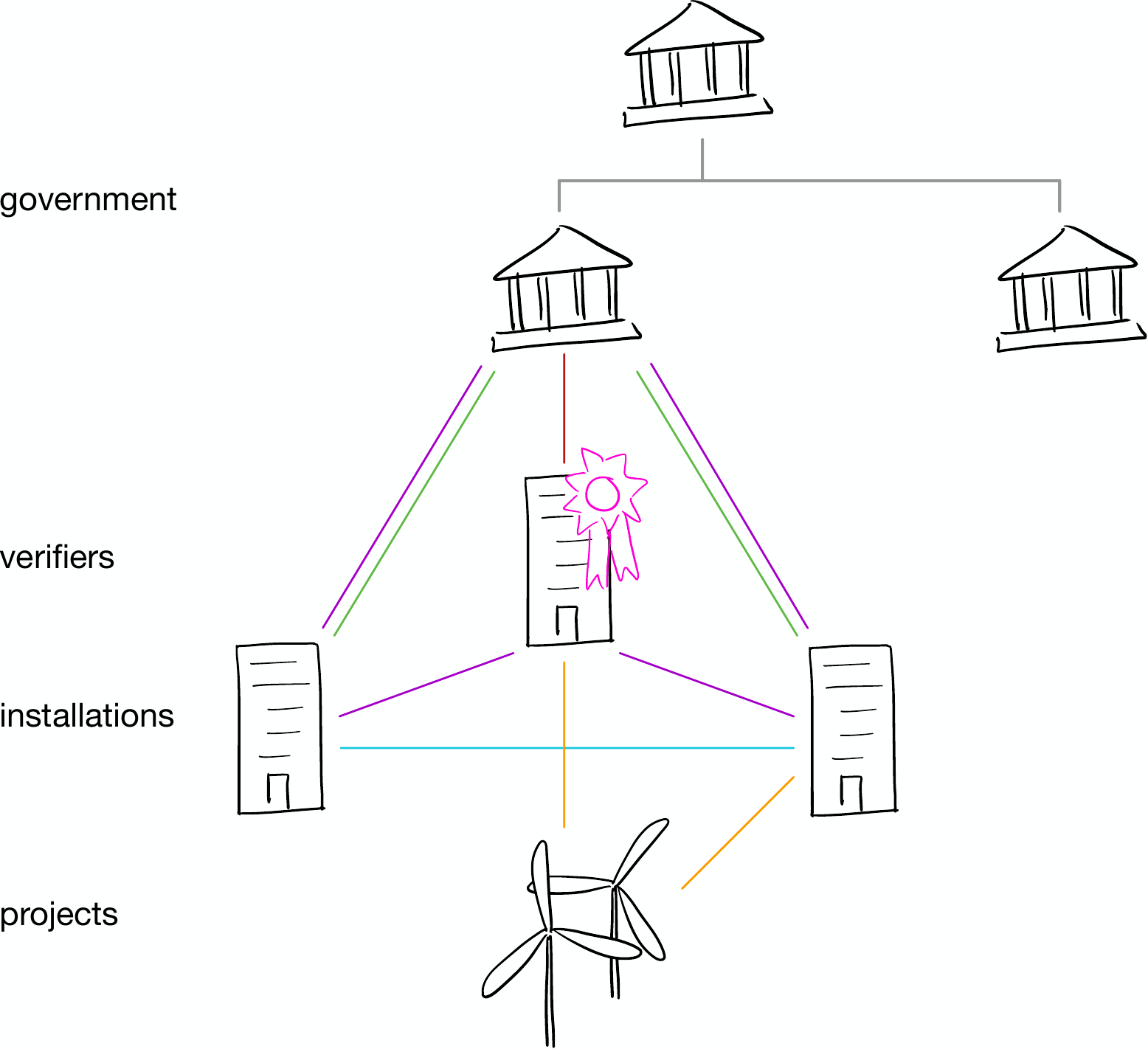}
    \caption[Overview of EU ETS]{Overview of EU ETS. Lines represent transactions between parties; the two layers of government represent the European Commission and member states.
    \begin{itemize}
        \item[\textcolor{green}{\myrule}] allowance issuance
        \item[\textcolor{purple}{\myrule}] verification and reporting of emissions and surrender of allowances and credits
        \item[\textcolor{lblue}{\myrule}] trading
        \item[\textcolor{orange}{\myrule}] credit issuance
        \item[\textcolor{dred}{\myrule}] oversight %through legislation, provision of verification frameworks and accreditation
    \end{itemize}}
    \label{fig:diagram-eu-ets}
\end{figure}

The EU ETS was launched in 2005 and has become the largest ETS to date, representing the majority of international emissions trading \cite{Braun2019ArePolicy}. Its coverage extends to over \num{11000} installations with significant energy usage as well as airlines operating in the EU, together representing about half of the EU's greenhouse gas (GHG) emissions\footnotemark{}  \cite{EuropeanCommissionEUETS}. A representative schematic of the different players and transactions of the EU ETS is presented in \cref{fig:diagram-eu-ets}.

\footnotetext{The GHGs covered by the EU ETS are carbon dioxide (\ch{CO2}), nitrous oxide (\ch{N2O}) and perfluorocarbons (PFCs).}

\runinhead{Tradeable instruments}
The EU ETS introduces a new tradeable instrument alongside the allowance unit: credits. Whilst allowances are issued by governments of member states through allocation or auction, credits are generated through emissions-reduction projects in other countries under Kyoto Protocol mechanisms. Any allowances or credits which are surplus to an installation's requirement to offset its emissions may be freely traded for profit \cite{EuropeanCommission2015EUHandbook}.

\runinhead{Impact}
Relative to a 2005 baseline, the EU ETS is expected to have reduced emissions by 21\% in 2020 and by 43\% in 2030, indicating that the underlying market mechanism is functioning as expected \cite{EuropeanCommissionEUETS}.

\subsection{Potential and suitability of blockchain}
Despite the successes highlighted previously, there still exist issues and challenges with both the EU ETS, and ETS more broadly, which this paper seeks to address. Specifically, we argue that the blockchain technology shows great potential to advance the state of the art in a number of key areas of ETS development.

\runinhead{Coverage}
Existing ETS are restricted in terms of geographical coverage, with large portions of the world currently lacking plans to implement ETS \cite{Braden2019BlockchainSystems}. A distributed scalable blockchain-based ETS solution could rapidly support new carbon markets. Article 6 of the 2018 Paris Agreement already provides a foundation for decentralised cooperative climate action; blockchain is expected to be a key technology to deliver these ambitions, particularly through future carbon markets \cite{AsianDevelopmentBank2018DecodingAgreement,Dinakaran2019HarnessingAgreement}.

\runinhead{Linkage}
A key challenge of existing ETS is that few systems are yet linked to each other. Although some ETS have previously implemented links, the process is complex and lengthy, as evidenced by the near decade-long process to link the Swiss and EU ETS \cite{FederalOfficefortheEnvironment2019LinkingSchemes,CounciloftheEU2019Linking2020}. For the UK, an exit from the EU---and consequently the EU ETS---may hinder attempts to meet future carbon budgets \cite{Hepburn2017ClimateBrexit}. In this context, an easily extensible linked ETS solution that can be rapidly deployed in new areas would be highly desirable.

An interlinked web of ETS would increase market liquidity and size \cite{InternationalCarbonActionPartnership2016OnSystems,Santikarn2018ASystems,PartnershipforMarketReadiness2016EmissionsImplementation,Dong2018BlockchainMarkets}, and reduce opacity relative to siloed systems. Transparently linking multiple ETS would increase the cost, and hence lower the chance, of fraudulently claiming credits from the same project in multiple systems (``double-counting'') \cite{Braden2019BlockchainSystems,Braden2019BlockchainInstruments,Dong2018BlockchainMarkets}.

\runinhead{Cost}
A (semi-)automated decentralised system can be expected to reduce overall transaction cost, especially if fixed costs are spread across a large network. Lower transaction cost will reduce barriers to entry, allowing coverage to be extended to smaller enterprises and less-developed geographies.

\runinhead{Trust}
Codified protocols in immutable smart contracts are tamperproof and thus blockchain ETS are expected to improve trust relative to existing systems \cite{Banerjee2018Re-EngineeringTechnology}. This could help maintain market confidence and integrity with linked ETS, for example if one ETS operates in a jurisdiction with increased risk of corruption \cite{Dong2018BlockchainMarkets}.

\runinhead{Transparency}
The shared, distributed nature of a blockchain system underpins transparency.
Address anonymity (or perhaps pseudonymity) with blockchain would allow transaction data to be made available in much greater detail, without compromising privacy or confidentiality concerning e.g.\ ETS players' trading positions. Compared to the EU ETS transaction log (EUTL), from which relatively little data is made available, increased scrutiny of public data could strengthen systems and reduce the risk of government corruption \cite{EuropeanCommission2015EUHandbook,Fuessler2018NavigatingAction,Braden2019BlockchainSystems,Braden2019BlockchainInstruments}.

\runinhead{Consensus and fault tolerance}
Consensus mechanisms provide a degree of fault tolerance that could mitigate the consequences of misbehaviour of network participants  \cite{Andoni2019BlockchainOpportunities,Aggarwal2019BlockchainOpportunities,Sikorski2017BlockchainMarket}. This is relevant particularly in the context of linking ETS, where many players will be connected to the system.

\subsection{Existing work}
Crucially, existing attempts to bring the benefits of blockchain to carbon trading have generally suffered from limited impact and short lifespan \cite{Fuessler2018NavigatingAction}, being largely predicated on the small voluntary carbon market. As such, any blockchain solution will require significant support from existing ETS regulators to ensure sufficient impetus for further growth and development. Once ``critical mass'' is achieved however, a solution can be expected to become self-sustaining; the contribution of regulators to facilitate access to the regulatory compliance market is likely to be a significant factor for success. It is also important to remember that blockchain is not a panacea, despite what may have been written about it in the press over the past years; the technology still suffers from important limitations \cite{Tucker2018WhatDo} (see discussion in \cref{sec:challenges-considerations}).

\Cref{tab:existing-work-overview} presents selected existing works related to the present discussion; whilst the concept of blockchain ETS has already been broadly discussed, there remains scope for additional work considering the practical implementation of such an ETS, which this paper seeks to address.

\begin{table}
    \caption{Overview of selected existing works}
    \label{tab:existing-work-overview}
    \centering
    
    \begin{tabularx}{\linewidth}{lX}
    \hline\noalign{\smallskip}
    Source & Description \\
    \noalign{\smallskip}\svhline\noalign{\smallskip}

    \multicolumn{2}{l}{\textit{Discussion papers}} \\

    \cite{AlKawasmi2015Bitcoin-BasedModel} & Presents a systems engineering approach to a decentralised emissions trading infrastructure. Reviews architecture (covering e.g. database type, credit issuance, existence of central authority etc.) of other carbon trading schemes. \\
    
    \cite{Braden2019BlockchainSystems} & Outlines blockchain potential for ETS, climate mitigation and climate finance applications in the specific context of Mexico, with discussion of potential implementation (technologies, costs, roadmap etc.). \\
    
    \cite{Braden2019BlockchainInstruments} & Provides an overview of blockchain potential, suitability and challenges for applications including ETS, MRV and climate finance. \\
    
    \cite{Dong2018BlockchainMarkets} & Describes current climate markets from a technological perspective and discusses improvements. Presents potential and suitability of blockchain without discussion of implementation. \\
    
    \cite{Fuessler2018NavigatingAction} & Argues for the suitability and potential of blockchain in achieving the commitments of the Paris Agreement and for climate action in general, with discussion of areas of required future work. \\
    
    \noalign{\smallskip}
    \multicolumn{2}{l}{\textit{Implementation work}} \\

    \cite{Eckert2019AMobility} & Discusses general suitability of blockchain for ETS applications, and presents a proof-of-concept implementation for a transportation-specifc ETS using Hyperledger Iroha. \\ 
    
    \cite{Imbault2017TheConsumption} & Proof-of-concept blockchain for green certificates (proof of electricity generation from renewable sources) in a microgrid electricity trading environment. Uses the Corda platform.\\
    
    \cite{Khaqqi2018IncorporatingApplication} & Proof-of-concept ETS implementation using `reputation points' to determine market access priority, thus aiming to tackle security issues identified with EU ETS. Includes detailed quantitative analysis of improvement relative to conventional systems. \\
    
    \cite{Liss2018BlockchainContracts} & Develops detailed proof-of-concept blockchain ETS based on EU ETS, using smart contracts on Ethereum. Discusses software development process (requirements, use cases) and system architecture (implementation) in detail. \\
    
    \noalign{\smallskip}\hline\noalign{\smallskip}
    \end{tabularx}
\end{table}
\section{Proposal}

\begin{figure}
    \sidecaption
    \includegraphics[width=75mm]{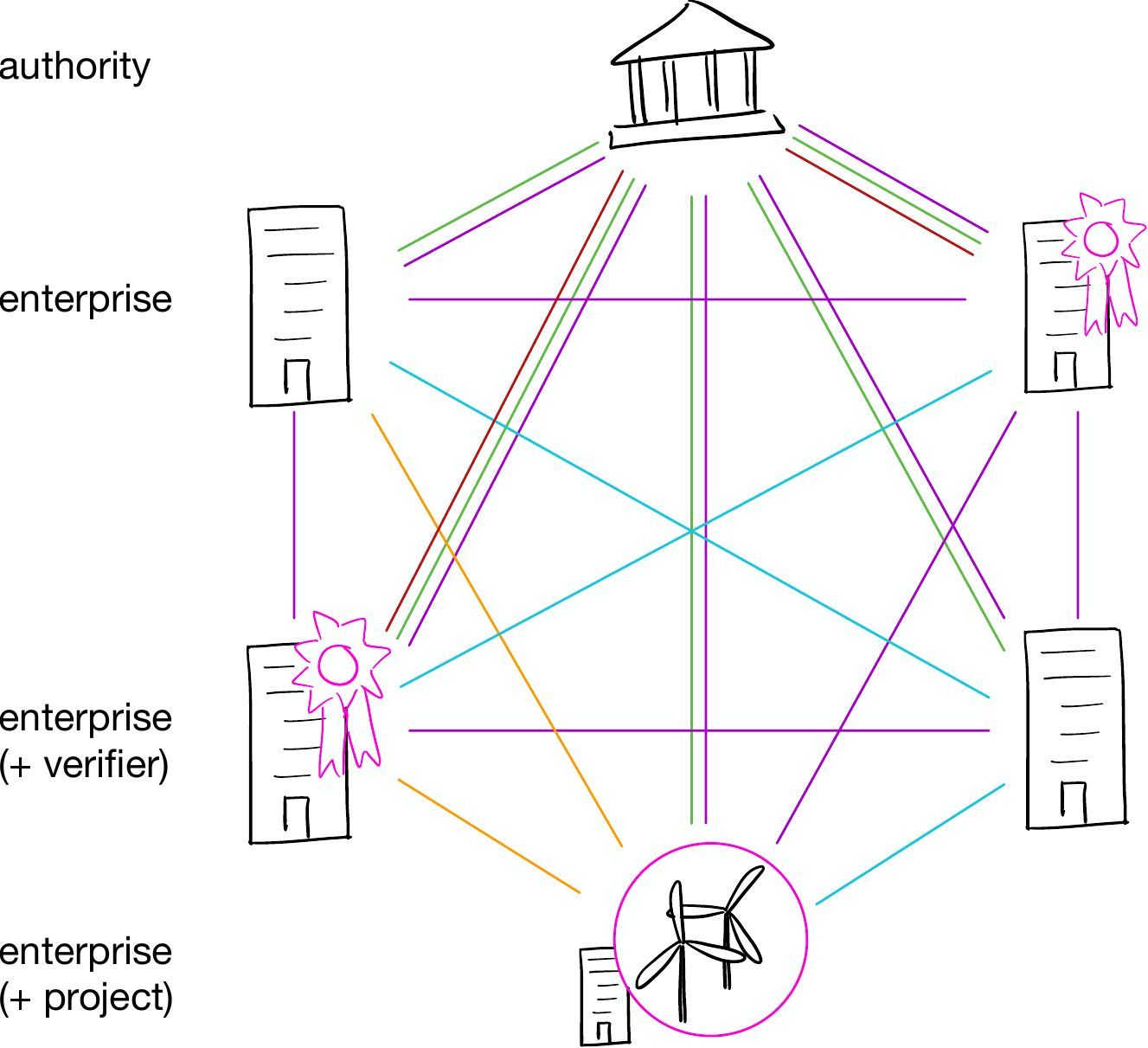}
    \caption[Schematic showing potential interactions in the outlined blockchain ETS]{Schematic showing potential interactions in the outlined blockchain ETS.
    \begin{itemize}
        \item[\textcolor{green}{\myrule}] allowance issuance
        \item[\textcolor{purple}{\myrule}] verification and reporting of emissions and surrender of allowances and credits
        \item[\textcolor{lblue}{\myrule}] trading
        \item[\textcolor{orange}{\myrule}] credit issuance
        \item[\textcolor{dred}{\myrule}] oversight
    \end{itemize}}
    \label{fig:diagram-blockchain-ets}
\end{figure}

Whilst it is tempting to develop a completely decentralised blockchain-based ETS, a more pragmatic approach might be to improve upon existing ETS frameworks. Thus, we propose a hybrid model combining some decentralisation whilst maintaining a role for trusted authorities, as outlined briefly in \cref{fig:diagram-blockchain-ets}. This does not preclude a future switch to a fully decentralised model, but it is expected to provide an easier transition to blockchain technology for existing ETS players, increasing the practical feasibility of the proposal.

\subsection{Taxonomy}
In this proposal the following definitions are used:

\begin{description}[organisation]
    \item[organisation] The simplest type of entity in the network, upon which other roles are built. An organisation provides a framework to manage common metadata required to interact with the blockchain (e.g. public/private key management).
    \item[authority] Governmental or supranational body, with legislative power over other authorities or enterprises within a certain jurisdiction.
    \item[enterprise] ``Legal person'' consisting of one or more installations or projects that uses the network to report and/or offset emissions. An enterprise may be mandated to participate by an authority, or access voluntarily and may carry verifier status.
    \item[installation] Physical source of GHG emissions---such as factories, manufacturing plants, offices---owned by one or more enterprises.
    \item[project] Emissions reduction scheme generating carbon credits under Kyoto Protocol mechanisms, owned by one or more enterprises.
    \item[verifier] Status awarded to an enterprise by an authority, allowing it to perform verification functions within the network.
\end{description}

\subsection{Tokens}
Two types of token are envisaged for network, closely mirroring current ETS book-keeping \cite{EuropeanCommission2015EUHandbook,Braden2019BlockchainSystems} (see \cref{app:rec}).

\begin{description}[allowance]
    \item[emission] \SI{1}{\tCe} verified GHG emissions.
    \item[permit] permit to emit \SI{1}{\tCe} that can represent both allowances issued by an authority, or credits granted by a verifier.
\end{description}

\subsection{Processes}

In this section, we present a basic outline of processes performed on the network, illustrated with smart contract pseudocode required for their execution.\footnotemark At the scale required of an international ETS, a custom-developed blockchain may be more appropriate; nevertheless, our approach provides a simple framework for presenting and debating a proof-of-concept.

\footnotetext{Our pseudocode is inspired by the \href{https://github.com/ethereum/solidity}{Solidity} language used to implement smart contracts on the Ethereum blockchain.}

\runinhead{Role change}
An authority can change the role of an enterprise, including promotion of an enterprise to verifier status or removal of an existing status.

\begin{algorithm}[H]
\caption{Role change}\label{alg:role-change}
\DontPrintSemicolon
\SetKwFunction{setRole}{setRole}

\Fn{\setRole{\KwType{address} sender, \KwType{address} target, \KwType{string} newRole}}{
    \KwRequire \KwVar{sender}.role = authority \KwAnd \KwVar{target}.role $\neq$ \KwVar{newRole} \tcp*{Sender must be authorised and request is a change}
    \KwVar{target}.role $\leftarrow$ \KwVar{newRole} \;
}
\end{algorithm}

\runinhead{Issuance of permit}
Emission permits can represent both allowances, issued by an authority, or credits, issued by a verifier.

An authority can mint permit tokens typically in an amount corresponding to the desired cap level. They may be issued through direct allocation or auction.

\begin{algorithm}[H]
\caption{Mint permit token}\label{alg:mint-allowance}
\DontPrintSemicolon
\SetKwFunction{mintAToken}{mintPermit}

\Fn{\mintAToken{\KwType{address} signer, \KwType{address} target, \KwType{uint256} amount}}{
    \KwRequire \KwVar{signer}.role = authority \label{line:mint-allowance:authority}
    \tcp*{Only authorities can mint permits}
    \KwVar{target}.balance[permit] += \KwVar{amount} \label{line:mint-allowance:receiver}\;
    market.balance[permit] += \KwVar{amount} \;
}
\end{algorithm}

Permit tokens can also represent credits granted by a verifier to an enterprise owning emission-reducing projects.

\begin{algorithm}[H]
\caption{Grant permit token}\label{alg:mint-credit}
\DontPrintSemicolon
\SetKwFunction{mintCToken}{grantPermit}
\SetKwFunction{hasProject}{hasProject}

\Fn{\mintCToken{\KwType{address} signer, \KwType{address} target, \KwType{uint256} amount}}{
    \KwRequire \hasProject{target} \KwAnd \KwVar{signer}.role = verifier
    \tcp*{Verifier ensures that the enterprise has a carbon-reducing project}
    \KwVar{target}.balance[permit] += \KwVar{amount} \;
    market.balance[permit] += \KwVar{amount} \;
}
\end{algorithm}

Both \texttt{mintPermit} and \texttt{grantPermit}
allow permit tokens to be issued ``out of thin air'', thus increasing the total circulating supply of the token in the market.

\runinhead{Issuance of emissions}
Emission tokens may be minted by any enterprise if a verifier co-signs the transaction as a true reflection of the enterprise's emissions.

\begin{algorithm}[H]
\caption{Mint emission token}\label{alg:mint-emission}
\DontPrintSemicolon
\SetKwFunction{mintEToken}{mintEmission}
\SetKwFunction{verifySignatures}{verifySignatures}

\Fn{\mintEToken{\KwType{address} sender, \KwType{address} signer, \KwType{uint256} amount}}{
    \KwRequire \KwVar{signer}.role = verifier \label{line:verisign}
    \tcp*{Must be signed by a verifier}
    \KwVar{sender}.balance[emission] += \KwVar{amount} \;
    market.balance[emission] += \KwVar{amount} \;
}
\end{algorithm}

\runinhead{Transfer tokens}
Permit tokens which represent emission allowances and credits may be freely transferred among network participants, who may choose to create derivative products such as swaps and options (as in the EU ETS \cite[p.71]{EuropeanCommission2015EUHandbook}) or to send tokens to an exchange.

\begin{algorithm}[H]
\caption{Transfer permit tokens}\label{alg:transfer-tokens}
\DontPrintSemicolon
\SetKwFunction{transferToken}{transferPermit}

\Fn{\transferToken{\KwType{address} sender, \KwType{address} target, \KwType{uint256} amount}}{
    \KwRequire \KwVar{amount} $\leq$ \KwVar{sender}.balance[permit] \;
    \tcp*{Must have enough token for request}
    \KwVar{sender}.balance[permit] --= \KwVar{amount}\;
    \KwVar{target}.balance[permit] += \KwVar{amount}\;
}
\end{algorithm}

\runinhead{Burn tokens}
Emissions tokens are burnt alongside an equal or greater number of allowance or credit tokens sent in the same transaction to a smart contract. This process also permits enterprises to voluntarily surrender excess allowance or credit tokens if they so choose (as is possible in the EU ETS \cite[p. 131]{EuropeanCommission2015EUHandbook}). Enterprises are forbidden from transacting with emissions tokens in any other way.

\begin{algorithm}[H]
\caption{Burn tokens}\label{alg:burn-tokens}
\DontPrintSemicolon
\SetKwFunction{burnToken}{burnToken}

\Fn{\burnToken{\KwType{address} sender, \KwType{uint256} amount}}{
    \KwRequire \KwVar{amount} $\leq$ \KwVar{sender}.balance[permit] \tcp*{Must have enough token}
    \uIf{\upshape \KwVar{sender}.balance[emission] $\geq$ \KwVar{amount}}{
        \tcp*{Only burning part of emission balance}
        \KwVar{sender}.balance[emission] --= \KwVar{amount} \;
    }
    \ElseIf{\upshape \KwVar{sender}.balance[emission] $<$ \KwVar{amount}}{
        \tcp*{Burning beyond emission balance (voluntary surrender)}
        \KwVar{sender}.balance[emission] = 0 \;
    }
    \KwVar{sender}.balance[permit] --= \KwVar{amount} \;
}
\end{algorithm}

\runinhead{Token exchange}
Organisations can freely trade their permit tokens with the authority. To ensure liquidity in the market and hence enhance the tradability of tokens, we can implement the Bancor protocol~\cite{Rosenfeld2017FormulasSystem,Hertzog2018BancorContracts} which automates price determination through a smart contract (see \cref{alg:tradeToken,alg:convertCash}, and Appendix~\ref{app:bancor}).

\begin{algorithm}[h]
\caption{Trade tokens}\label{alg:tradeToken}
\DontPrintSemicolon
\SetKwFunction{tradeToken}{tradeToken}

\Fn{\tradeToken{\KwType{address} sender, \KwType{int256} amount}}{
    supply $\leftarrow$ market.balance[permit] \;
    cashAmount $\leftarrow$  reserve * ((1 + \KwVar{amount}/supply )\textasciicircum (1/fraction) -- 1) \;
    \tcp*{Based on \cref{eq:ttoe} in Appendix}
    \uIf{amount $> 0$}{
    \KwRequire cashAmount $\leq$ \KwVar{sender}.cash
    \tcp*{Must have cash to spend}
    \KwVar{sender}.balance[permit] += \KwVar{amount} \;
    market.balance[permit] += \KwVar{amount} \;
    \KwVar{sender}.cash --= cashAmount \;
    reserve += cashAmount \;
    }
    \ElseIf{amount $<= 0$}{
    \KwRequire \KwVar{amount} $\leq$ \KwVar{sender}.balance[permit]
    \tcp*{Must have token to sell}
    \KwVar{sender}.balance[permit] += \KwVar{amount} \;
    market.balance[permit] += \KwVar{amount} \;
    \KwVar{sender}.cash += cashAmount \;
    reserve --= cashAmount \;
    }
}
\end{algorithm}

\begin{algorithm}[h]
\caption{Convert Cash}\label{alg:convertCash}
\DontPrintSemicolon
\SetKwFunction{convertCash}{convertCash}

\Fn{\convertCash{\KwType{address} sender, \KwType{int256} amount}}{
supply $\leftarrow$ market.balance[permit] \;
    tokenAmount $\leftarrow$ supply * ((\KwVar{amount}/ reserve + 1)\textasciicircum fraction $- 1)$ \;
    \tcp*{Based on \cref{eq:etot} in Appendix}
    \uIf{\KwVar{amount} $> 0$}{
    \KwRequire \KwVar{amount} $\leq$ \KwVar{sender}.cash
    \tcp*{Must have cash to spend}
    \KwVar{sender}.balance[permit] += tokenAmount \;
    market.balance[permit] += tokenAmount \;
    \KwVar{sender}.cash --= \KwVar{amount} \;
    reserve += \KwVar{amount} \;
    }
    \ElseIf{amount $<= 0$}{
    \KwRequire tokenAmount $\leq$ \KwVar{sender}.balance[permit]
    \tcp*{Must have token to sell}
    \KwVar{sender}.balance[permit] --= tokenAmount \;
    market.balance[permit] --= tokenAmount \;
    \KwVar{sender}.cash += \KwVar{amount} \;
    reserve --= \KwVar{amount} \;
    }
}
\end{algorithm}

\subsection{Market adjustment}
With the development of technology, the cost of emissions reduction will decrease over time. As a result, the supply of surplus allowances and credits will increase whilst demand for them decreases, driving the price of tokens down. Thus it may become cheaper for firms to use credits to offset their emissions rather than reducing the emissions directly.

\begin{figure}
    \includegraphics[width=0.8\linewidth,trim={10 43 10 10}, clip]{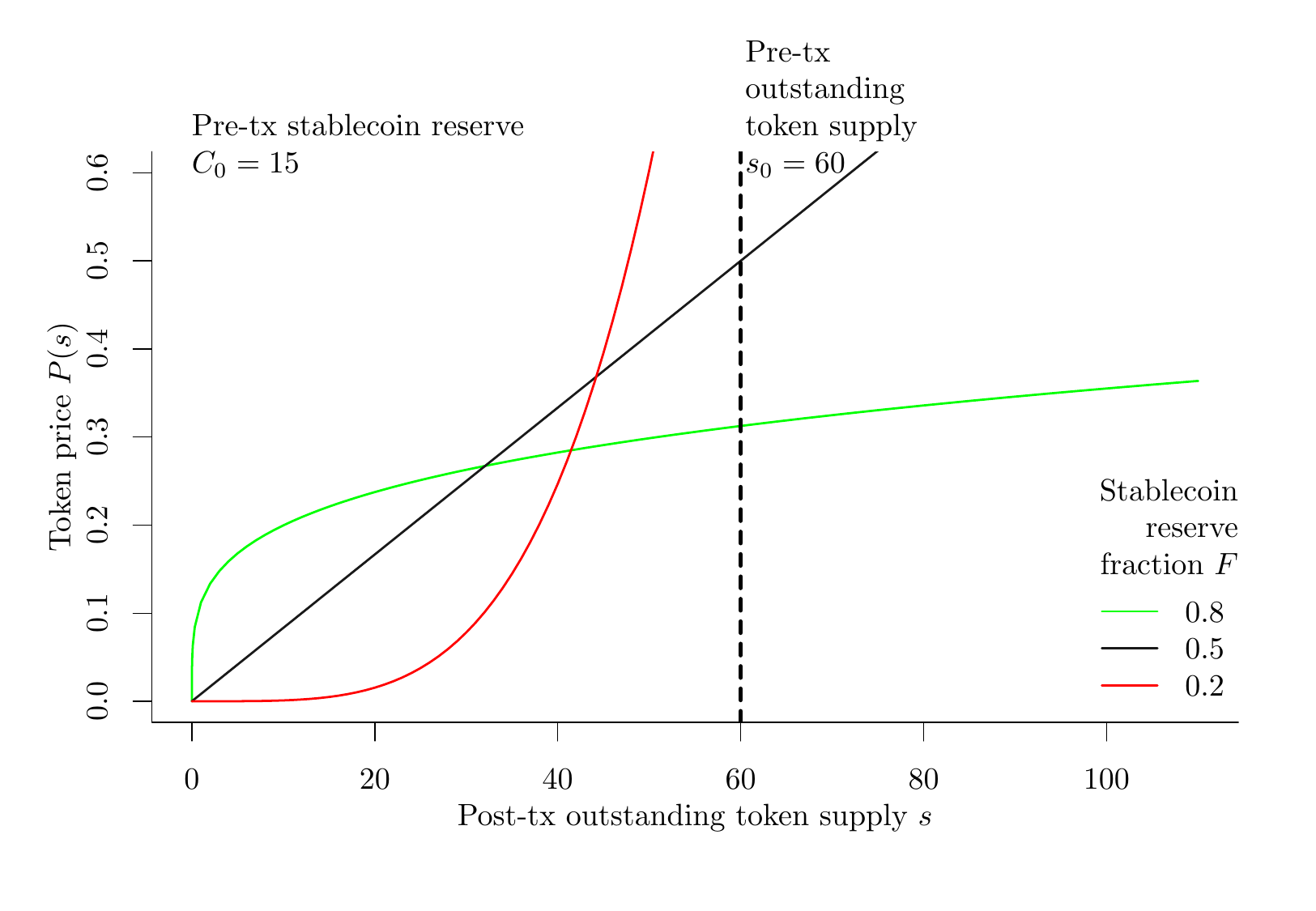}
    \caption{Token price as a function of supply as specified in (\ref{eq:ps}).}
    \label{fig:tokenprice}
\end{figure}

This is when the authority comes into play to steer the market. In addition to having the power to change the cap level and so restrict the supply of allowances, the price of tokens can also be adjusted through the exchange. From (\ref{eq:ps}) and its illustration presented in \cref{fig:tokenprice}, it is clear that the tokens' market price can be raised in two ways:
\begin{itemize}
    \item Reducing reserve fraction $F$, allowing for cash to be spent from the exchange and thus enhancing its purchasing power;
    \item Increasing the total stablecoin reserve $C_0$, thus enhancing the purchasing power of the exchange;
\end{itemize}
Naturally, to reduce the token price, the authority simply needs to do the opposite.

\subsection{Carbon bookkeeping on and off blockchain}
\label{app:rec}
We demonstrate the compatibility of our proposed blockchain-based model with the existing ETS frameworks through an illustrative example. In principle, blockchain is underpinned by time-honoured bookkeeping mechanisms including TEA (Triple Entry Accounting) and REA (Resources-Events-Agents) \cite{Ibanez2020REASystems,Ibanez2020TheAccounting}. Therefore, the compatibility between the conventional and the newly proposed ETS record-keeping framework is expected to an extent.

In our illustrative example, we assume that:
\begin{enumerate}
    \item On January 1, 2020, the market value for of one permit token was 20 \euro.
      \begin{itemize}
        \item  Authority $\mathscr{A}$ allocated Enterprise $\mathscr{E}$ with allowances for \SI{100}{\tCe} by issuing $\mathscr{E}$ 100 permit tokens.
        \item  Verifier $\mathscr{V}$ approved Enterprise $\mathscr{E}$'s carbon-reducing project in a developing country and granted  $\mathscr{E}$ with credits for \SI{40}{\tCe} by issuing $\mathscr{E}$ 40 permit tokens.
        \item  Enterprise $\mathscr{E}$ transferred 10 permit tokens to Enterprise $\mathscr{F}$.
      \end{itemize}
    \item On June 30, 2020, the market value of one permit token increased to 24 \euro.
      \begin{itemize}
        \item  Enterprise $\mathscr{E}$ recorded \SI{55}{\tCe} emissions during January and June.
        \item  Enterprise $\mathscr{E}$ cashed out 240 \euro\ by selling 10 tokens to the market.
      \end{itemize}
    \item On December 31, 2020, the market value of one permit token decreased to 22 \euro.
      \begin{itemize}
        \item  Enterprise $\mathscr{E}$ recorded \SI{70}{\tCe} emissions during June and December.
        \item  Enterprise $\mathscr{E}$ bought 5 permit tokens from the market to cover the emissions.
        \item  Enterprise $\mathscr{E}$ surrenders 125 permit tokens to offset the total emissions of \SI{125}{\tCe} during year 2020.
      \end{itemize}
\end{enumerate}

In \cref{tab:journal}, we juxtapose smart contract execution with double-entry journalisation to show the correspondence of the two systems. We use the fair value method to record ``Emission permit'', following \cite{Ratnatunga2011TheManagement, Oker2017ReportingStandards}.

\begin{table}[H]
  \centering
  \caption{Carbon accounting with smart contract execution and journalisation}
    \begin{tabularx}{\linewidth}{X|lXrr}
    \toprule
    \emph{Smart contract execution} & \multicolumn{2}{l}{\emph{Journalisation}} &  \emph{Dr.}  &  \emph{Cr.}  \\
    \midrule
    \texttt{mintPermit}($\mathscr{A}$, $\mathscr{E}$, 100) & \textbf{Asset} & Emission permit-- Allowances &   2,000   &  \\
          & \textbf{Liability} & \qquad Deferred income &       &   2,000   \\
    \midrule
    \texttt{grantPermit}($\mathscr{V}$, $\mathscr{E}$, 40) & \textbf{Asset} & Emission permit -- Credits &      800   &  \\
          & \textbf{Liability} & \qquad Deferred income &       &      800   \\
    \midrule
    \texttt{transferPermit}($\mathscr{E}$, $\mathscr{F}$, 10) & \textbf{Asset} & Deferred income &      200   &  \\
          & \textbf{Liability} & \qquad Emission rights &       &      200   \\
    \midrule
          & \textbf{Asset} & Emission permit &      480   &  \\
          & \textbf{Equity} & \qquad Gain on revaluation &       &      480   \\
    \midrule
    \texttt{mintEmission}($\mathscr{V}$, $\mathscr{E}$, 55) & \textbf{Liability} & Deferred income &   1,100   &  \\
          & \textbf{Equity} & \qquad Income &       &   1,100   \\
    & \textbf{Equity} & Expenses -- Emissions &      1,320   &  \\
          & \textbf{Liability} & \qquad Permit surrenderable &       &     1,320   \\
    \midrule
    \texttt{convertCash}($\mathscr{E}$, -240) & \textbf{Asset} & Cash  &      240   &  \\
          & \textbf{Asset} & \qquad Emission permit &       &      240   \\
          & \textbf{Equity} & Deferred income &      200   &  \\
          & \textbf{Equity} & \qquad Income &       &      200   \\
    \midrule
          & \textbf{Equity} & Loss on revaluation &      240   &  \\
          & \textbf{Asset} & \qquad Emission permit &       &      240   \\
    \midrule
    \texttt{mintEmission}($\mathscr{V}$, $\mathscr{E}$, 70) & \textbf{Liability} & Deferred income &   1,300   &  \\
          & \textbf{Equity} & \qquad Income &       &   1,300   \\
    & \textbf{Equity} & Expenses -- Emissions &      1,430   &  \\
          & \textbf{Liability} & \qquad Permit surrenderable &       &     1,430    \\
    \midrule
    \texttt{tradeToken}($\mathscr{E}$, 5) & \textbf{Asset} & Emission permit &      110   &  \\
          & \textbf{Asset} & \qquad Cash  &       &      110   \\
    \midrule
    \texttt{burnToken}($\mathscr{E}$, 125) & \textbf{Liability} & Permit surrenderable &   2,750   &  \\
          & \textbf{Asset} & \qquad Emission permit &       &   2,750   \\
    \bottomrule
    \end{tabularx}%
  \label{tab:journal}%
\end{table}%
\section{Further challenges and considerations}
\label{sec:challenges-considerations}

\runinhead{Implementation}
The specific platform chosen to host a blockchain ETS is a particularly important consideration. In \cite{Braden2019BlockchainInstruments}, Bitcoin, Ethereum, Hyperledger Fabric and EOS are evaluated for climate policy applications, considering programmability, operating cost, security and usability, with the conclusion that Ethereum and Hyperledger Fabric are found to be the most promising platforms today. Similarly, \cite{Liss2018BlockchainContracts} considers Ethereum and Hyperledger Fabric as strong candidate frameworks for ETS implementation, noting important distinctions between the two: Ethereum is by default public and permissionless; Hyperledger Fabric is private and permissioned. A more complete discussion of many other platforms can be found in  \cite{Aggarwal2019BlockchainOpportunities}.

As discussed previously, both developing a derivative blockchain solution (such as using Hyperledger Fabric) and developing an entirely custom implementation should be considered in finding an approach to host ETS at large scale. The relative benefit of building upon an established system (e.g. pre-existing audited and/or open-source code) must be weighed against the degree of customisation desired. Further, should differing implementations be developed by governments or organisations, standardisation could still enable interoperability \cite{Deshpande2017UnderstandingStandards}.

\runinhead{Governance and trust}
Since the ``the allocation of allowances, the opening and closing of ETS registry accounts or the recognition of offset credits are ... [still] sovereign tasks of the government'', a comprehensive carbon network would by default involve governments as central authorities \cite{Braden2019BlockchainSystems}. The initial delegation of authority in a permissioned blockchain ETS requires participants to trust the authority establishing the network and thus the integrity of the tokens issued. With ETS linkage that connects different states and regions, it cannot be guaranteed that every participant will trust all authorities equally, imposing the need for an on-chain governance design that ensures the integrity of authorities. Importantly, whilst smart contracts may be ideally suited to the rigorous application of defined rules, these rules must first be developed in collaboration with stakeholders \cite{Dong2018BlockchainMarkets}.

Further, a potential future shift towards a decentralised blockchain without explicit governmental oversight presents a significant complication: should there be trust asymmetries between players, the fungibility of tokens issued by different entities will be challenged and could lead to fragmentation of the network. One solution could be standardisation \cite{Deshpande2017UnderstandingStandards}.

\runinhead{Enforcement}
Current ETS utilise legislation to compel enterprises to participate. Whilst a voluntary carbon market does also exist, it is significantly smaller than the regulatory compliance market \cite{Seeberg-Elverfeldt20102Work}. In a completely decentralised international ETS, it is less clear what would motivate participants. Additionally, defining how criminal activity on the network would be deterred is challenging, potentially requiring a supranational enforcement body to maintain network integrity; indeed, ``new governance systems will be needed to ensure market and environmental integrity in a peer-to-peer environment'' \cite{Dong2018BlockchainMarkets}.

\runinhead{Measurement, reporting and verification (MRV)}
A critical issue with any blockchain solution is its interface with the real world \cite{Tucker2018WhatDo}; the maxim ``garbage in, garbage out'' aptly illustrates the consequences of poor input data. The verification and accreditation processes in the EU ETS are complex and potentially burdensome \cite{EuropeanCommission2015EUHandbook}. Moving beyond the model of trusted verifiers to a truly decentralised approach will require significant effort to develop alternative MRV methodologies.

The internet of things (IoT) will enable a universally trusted mechanism for MRV of real-world data, by automating data flows and processes \cite{Dong2018BlockchainMarkets,Fuessler2018NavigatingAction}. IoT technology is expected to reduce the cost and time requirement of MRV, whilst enhancing trust through increased reliability and the accessibility of audited code. Real-time sensing will enable a faster compliance cycle than the current yearly process in the EU ETS \cite{Braden2019BlockchainSystems,Fuessler2018NavigatingAction}. Increased trading activity through more frequent reporting and compliance will enhance market liquidity. Additionally, diverse data sources such as earth observation satellites will enable stronger verification of reported emissions or emissions reductions.

\runinhead{Performance}
It is well known that blockchain networks can have extremely poor performance relative to simple databases, especially in electricity usage \cite{Dong2018BlockchainMarkets,Aitzhan2018SecurityStreams}. One study has estimated that the global Bitcoin network consumes approximately as much power as the country of Ireland, and forecasts this consumption more than tripling in future \cite{deVries2018BitcoinsProblem}. Various alternative consensus mechanisms have been proposed to improve performance and reduce the environmental impact of the infrastructure itself \cite{Andoni2019BlockchainOpportunities, Perez2020WeTransactions}.
\section{Conclusion}

Having investigated the potential applicability of blockchain technology to carbon trading on ETS, we can conclude that although there is potential for blockchain to enhance the impact and reach of current ETS in a number of ways, significant barriers remain, limiting the applicability of the technology today. A basic outline of a permissioned blockchain solution largely replicating today's EU ETS has been presented as a viable transitional first step towards the development of a fully-decentralised blockchain ETS, which could significantly accelerate the deployment of this important emissions reduction tool worldwide. We conclude however that significant legislative and legal barriers remain to be overcome for a decentralised blockchain ETS to realise its full potential on implementation. 

\section*{Appendix}
\renewcommand{\thesubsection}{\Alph{subsection}}
\subsection{Bancor algorithm for token exchange}
\label{app:bancor}

As demonstrated with \cref{alg:tradeToken,alg:convertCash}, the Bancor exchange protocol~\cite{Rosenfeld2017FormulasSystem,Hertzog2018BancorContracts} ensures constant tradability of a token, as it prices a token algorithmically, as opposed to through matching a buyer and a seller.  We use the notation outlined in \cref{tab:token-exchange-notation} to explain the protocol.

For demonstration purposes, we assume that the medium of exchange is a stablecoin, measured in \euro, that circulates on the same blockchain as the permit tokens.

\begin{table}
\caption{Mathematical notation for token exchange}
\label{tab:token-exchange-notation}
\begin{tabularx}{\textwidth}{lXl}
    \noalign{\smallskip}\hline\noalign{\smallskip}
    Notation & Definition                     & Unit     \\
    \noalign{\smallskip}\svhline\noalign{\smallskip}
    
    \multicolumn{3}{l}{\textit{Preset hyperparameters, occasionally adjusted}}                      \\
    $F$     & Constant fraction of stablecoin reserve   & ---               \\
    \noalign{\smallskip}
    
    \multicolumn{3}{l}{\textit{Input variables}}                      \\
    $s_0$   & Pre-transaction outstanding token supply                    & tokens            \\
    $C_0$   & Pre-transaction stablecoin reserve                   & \euro/token       \\
        $e$     & Tokens bought (negative when sold)                & tokens            \\
    $t$     & Stablecoins spent (negative when received)              & \euro             \\
    \noalign{\smallskip}
    
    \multicolumn{3}{l}{\textit{Output variables}}                      \\
    $s$     & Post-transaction token supply                      & tokens            \\
    $P(.)$     & Post-transaction token price, dependent on token supply $s$ \                                  & \euro/token       \\
    $C(.)$     & Post-transaction stablecoin reserve, dependent on token supply $s$                                  & \euro             \\
    
    \noalign{\smallskip}\hline\noalign{\smallskip}
\end{tabularx}
\end{table}

It holds that, the stablecoin reserve $C$ (in \euro), always equals a fraction, predetermined as $F$, of the product of token price $P$ (in \euro/token) and outstanding token supply $s$ (in tokens). That is, the following equation is always true:

\begin{align}
C(s) &\equiv F\, s\, P(s) \label{eq:coinres}   \\
\shortintertext{Taking the derivative with respect to $s$ on both sides:}
\frac{\diff C(s)}{\diff s} &\equiv F\left[P(s)+s\frac{\diff P(s)}{\diff s}\right]
\label{eq:dd}
\end{align}

There exists another relationship between $C$, $P$ and $s$: if one buys from the exchange an infinitesimal amount of tokens, $\diff s$, when the outstanding token supply is $s$, then the unit token price at purchase would be $P(s)$. The exchange receives stablecoins and thus its reserve increases according to:

\begin{align}
\diff C(s)      &= P(s)\, \diff s \nonumber\\
\shortintertext{Rearranging:}
P(s)    &= \frac{\diff C(s)}{\diff s} \label{eq:price}
\end{align}

Combining (\ref{eq:dd}) and (\ref{eq:price}), we can derive price $P(.)$ as a function of $s$ as follows, which ultimately leads to (\ref{eq:ps}):

\begin{align}
P(s) & =F\,\left[P(s)+s\frac{\diff P(s)}{\diff s}\right]
\nonumber \\
\frac{\diff P(s)}{P(s)} & =\left(\frac{1}{F}-1\right)\frac{\diff s}{s}
\nonumber \\
\shortintertext{Integrating over $s \in (s_0, s)$:}
\int_{x=P(s_0)}^{P(s)}\frac{\diff x}{x} & =\left(\frac{1}{F}-1\right) \int_{y=s_0}^s\frac{\diff y}{y}
\nonumber \\
\ln P(s) - \ln \frac{C_0}{F\,s_0} & =\left(\frac{1}{F}-1\right)(\ln s - \ln s_0)
\nonumber \\
% \ln\frac{P}{P_0} & =\ln\Big(\frac{s}{s_0}\Big)^{\frac{1}{F}-1}
% \nonumber \\
P(s) & = \frac{C_0}{F\,s} \sqrt[F]{\frac{s}{s_0}}
\label{eq:ps}
\end{align}

Plugging (\ref{eq:ps}) into (\ref{eq:coinres}), we can derive the exchange's stablecoin reserve $C(.)$ as a function of $s$:

\begin{equation}
 C(s)=F\, s\, \frac{C_0}{F\,s} \sqrt[F]{\frac{s}{s_0}} = 
C_0\, \sqrt[F]{\frac{s}{s_0}}
\label{eq:cs}
\end{equation}

Assume one spends $t$ amount of stablecoins in exchange for $e$ amount of tokens when the outstanding token supply equal $s_0$. After the purchase, the outstanding token supply becomes $s_0+e$, while the stablecoin reserve increases by $t$, i.e.,

\begin{align}
t + C_0 = C(s_0+e)
\overset{\text{according to } (\ref{eq:cs})}{=}
C_0\, \sqrt[F]{\frac{s_0+e}{s_0}}
=
C_0\, \sqrt[F]{1+\frac{e}{s_0}}
\label{eq:prelim}
\end{align}

Rearranging (\ref{eq:prelim}), we get: 

\begin{itemize}
    \item the amount of stablecoins one must pay, denoted by $t$, based on the amount of tokens one wishes to receive, denoted by $e$, and the outstanding token supply $s$ (\cref{alg:tradeToken}),
    \begin{equation}
    \label{eq:ttoe}
    t =  C_0 \Bigg(\sqrt[F]{1+\frac{e}{s_0}}-1\Bigg)
    \end{equation}
    \item the amount of tokens one will receive, denoted by $e$, based on the amount of stablecoins one is willing to pay, denoted by $t$, and the outstanding token supply $s$ (\cref{alg:convertCash}).
    \begin{equation}
    \label{eq:etot}
    e = s_0\,\left[\left(\frac{t}{C_0} + 1\right)^F - 1\right]
    \end{equation}
\end{itemize}

% \newpage

\bibliographystyle{spmpsci}
\bibliography{references}

\end{document}